\documentstyle[mprocl,epsf]{article}

\def\Journal#1#2#3#4{{#1} {\bf #2}, #3 (#4)}

\def\JSP{\em J. Stat. Phys.}

\def\be{\begin{equation}}
\def\ee{\end{equation}}
\def\bea{\begin{eqnarray}}
\def\eea{\end{eqnarray}}

\begin{document}

\title{PARALLEL ALGORITHMS ON THE ASTRA SIMD MACHINE}

\author{G. \'ODOR}

\address{Research Institute for Materials Science, P.O.Box 49,
H-1525 Budapest, Hungary}

\author{F. ROHRBACH}

\address{CERN, CH-1211 Geneva 23, Switzerland }

\author{G. VESZTERGOMBI}

\address{Research Institute for Particle Physics, P.O.Box 49,
H-1525 Budapest, Hungary}

\author{G. VARGA, F. TATRAI}

\address{Technical University of Budapest, Department of Chemical Technology
H-1521 Budapest, Hungary}

\maketitle\abstracts{In view of the tremendous computing power jump of modern
RISC processors the interest in parallel computing seems to be thinning out.
Why   use a complicated system of parallel processors, if the problem
can be solved by a single powerful micro-chip? It is a general law, however,
that  exponential growth  will always end by some kind of a saturation, and
then   parallelism will again become   a hot topic.
We try to prepare ourselves for  this eventuality.
The MPPC project   started in 1990 in the heydeys of parallelism and
produced four ASTRA machines (presented at CHEP'92) with 4k processors (which
are expandable to 16k) based on `yesterday's chip-technology' (chip presented
at CHEP'91).
These machines now provide excellent test-beds for algorithmic
developments in a complete, real environment. We are developing for example
fast-pattern  recognition algorithms which
could be used in high-energy physics experiments at the LHC (planned to be
operational after 2004 at CERN) for
triggering and data reduction. The basic feature of our ASP (Associative String
Processor) approach is to use extremely simple (thus very cheap) processor
elements but in huge quantities (up to millions of processors) connected
together by a very simple string-like communication chain.
In this paper we present powerful algorithms based on this architecture
indicating the performance perspectives if the hardware (i.e. chip
fabrication) quality reaches `present or even future technology levels'.
}

\section{Introduction}

In the last five years interesting developments have taken place:  the
spectacular success of RISC machines, the diminishing interest in
 parallel machines, and
 the simultaneous disappearance of
mainframes. The relative cheapness of PCs giving workstation performance
together with
 extensive networking created a situation
that can be regarded as the MIMD version of parallel computing. Owing to the
expected increase of the bandwidth of network communication lines, these loose
MIMD systems will become more and more tightly connected. Therefore, if one
wants to speak about `real' parallel computing one should define it in a
restricted sense, close to the SIMD architecture. Here we  concentrate
on parallel computing based on simple processing elements but in enormous
quantities, which are so cheap that even for millions of them the price is
competitive. The motivation for studying  these
systems is two-fold. On the one hand,  the seemingly  exponential growth of
RISC performance, sometime in the future,  will
reach saturation. The interest for greater performance will promote again the
ideas of `tight' parallelism. On the other
hand, the biological systems  give us a convincing lesson about the
effectiveness of this type of system
based on billions of cheap neurons.

\subsection{Present state}\label{subsec:present}

At  present, the complexity of SIMD parallel systems lies typically  on
the level of a few thousands processors, providing general performance not
much above the  RISC machines in a small cluster.
A typical example of this class of system
is the ASTRA machine which was developed at CERN
by the MPPC collaboration.
The Associative String Processor (ASP) was devised by ASPEX Microsystems Ltd.
The first implementation of the machine,
 {\bf A}SP {\bf S}ystem {\bf T}est-bed for {\bf R}esearch
and  {\bf A}pplication (ASTRA)
 was installed in 1992 at CERN, Orsay, Saclay, and Brunel (ASPEX),
each with $4096$ processors. The results of the MPPC project were summarized
in the Final Report\cite{aspfin}. These machines (which are extendible to 16K)
based on `yesterday's chip-technology' provide excellent test-beds for
algorithmic developments in a complete, real environment. Because of this, in
1995 they are no longer competitive with the
latest generation of sequential machines.
However, by scaling up and extrapolation,
 one can gain experience  with them  for
future real massively parallel systems relying on forthcoming technologies.

\subsection{Massively parallel perspectives}\label{subsec:pers}

The new generation of SIMD machines will be competitive only if they are based
on extremely cheap elements. Among the Si-based components of a computer, the
cheapest ones are the memory chips where great demand justified the
application of the most advanced technology for mass production. NEC, the
Japanese electronics company, has reached another milestone\cite{NewSci}
in technology's steady march towards smaller, faster, and powerful computers.
This firm announced the production of the first prototype of a Gigabit dynamic
random access memory (DRAM) chip at the International Solid State Circuits
Conference in San Francisco.
NEC expects it will take more than three years and a further \$1.5 billion
to produce samples for computer manufacturers.
The aim of this paper is to demonstrate that,
 after rephrasing the ASP philosophy
as basically memory-oriented architecture,
 one can capitalize on the memory
technology in order
to propose cheap SIMD machines with millions of processing elements.
The versatility of this architecture will be illustrated by three algorithms
developed on the existing ASTRA machine.

\section{Boolean matrix exponentiation}

For us,  the problem of Boolean matrix exponentiation   originated
from the search for  strongly coupled components  in a set of equations
describing the control process\break\hfill\newpage\noindent of large chemical
plants. It implies
 the calculation of
$L$-th power of the $A$ adjacency matrix,  a well-known procedure in
mathematics:
\begin{equation}
  \label{adj}
  R^L=(I+A)^L~.
\end{equation}
If one has $n\times n$ matrix and $L=n$, then the processing time will be
${\cal O}(n^4)$ using the straightforward sequential algorithm. We propose
a parallel algorithm  which solves this problem in ${\cal O}(n\times\log n)$
time,  and represents a speed-up factor of a few million for $n > 1000$.
The algorithm was tested for small, $n=8,16,32,64$ cases on the existing ASTRA
machine. However, the real interest lies  in the large $n$ values. In
principle,
 the
program is completely identical for all $n$, if the number of processors is
scaled-up to $n^2$, and the individual memory size of each processors scales
with $n$. In ASTRA we have $64^2=4096$ processors each having $64$ bits
memory. We found that for this procedure
and in general for matrix calculations one does not need all the
associativity features which are
built into the ASP architecture, and therefore  propose  a reduced
ASP model to demonstrate the performance of our algorithm for large $n$ values.

\subsection{Bit matrix machine}\label{subsec:bitm}

In a memory chip the bits are ordered along a matrix which are selected
by some addressing logics. In our bit matrix machine model we propose to
replace this addressing scheme on one side of the memory matrix
by a string of one-bit processors with a
content addressable Activity Register. More precisely, the memory is
organized in the so-called `orthogonal-way' \cite{sopron}. It contains $m=n^2$
words of  $n$-bits length.
\begin{figure}[htbp]
  \begin{center}
    \leavevmode
         \epsfxsize=8.0cm
         \epsfbox{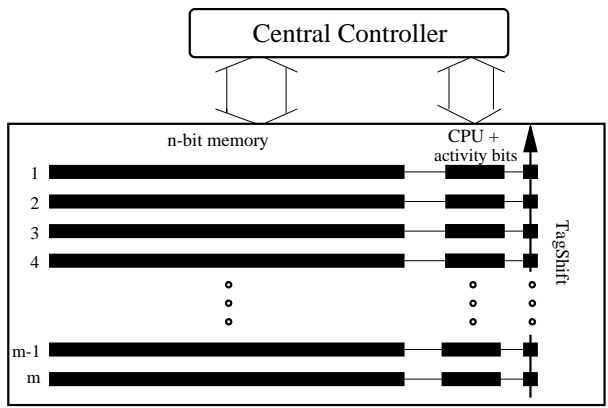}
         \vspace*{0.01cm}
    \caption{Bit matrix machine model}
  \end{center}
\end{figure}
\noindent
Each line contains an $n$-bit Data Register,
a $1$-bit CPU, a $6$-bit Activity Register, and the Tagging Bit, which
provides   communication amongst the processors by the
standard ASP TAGSHIFT operation\cite{orth}.
The total string can be subdivided into $n$ independent substring segments
(Fig. 1).

Although the `processor' part of the chip requires more gate circuits per bit,
 it uses (for large $n$ values)  a relatively much smaller portion of the
Si-area
than the `memory' part. Thus, as a whole its production  will require basically

the usual chip technology.

The READ/WRITE operation will be organized by selecting the
{\it horizontal} lines of the bit matrix Data Register.
The parallel processing
will take place along the {\it vertical} lines.

\subsection{Matrix algorithms}

The basic concepts of the bit-matrix machine are:
\begin{itemize}
\item The Memory (which is cheap) contains $n$-fold copies of the
bit-matrix in $n^3$ bits. This provides {\it column-wise} vectorization.
\item  The Activity-Register (which is relatively expensive) contains
the bit-matrix in $n^2$ bits providing {\it row-wise} vectorization.
\item The $n^2$ processors assure $n^2$-fold parallelism in the $a_{ij}$
matrix element level and $n$-fold parallelism on row/column level.
This provides execution times ${\cal O}(1)$ for addition,
${\cal O}(n)$ for multiplication and transposition, and ${\cal O}[n\cdot\log
(n)]$
for the calculation of $n$-th power.
\item The key element is the dispersed row representation
from which all the other matrix representations are reachable in
${\cal O}(n)$ steps.
\end{itemize}
If the technological progress  continues at the present rate, a chip
with $n = 30~000$ processor, each equipped with $30~000$ bits of memory, should
not be much larger than the present $1$ Gigabit memory-chip prototype
of NEC, which makes our model machine concept closer to reality.

\section{Simulation of damage spreading}

\subsection{Measuring the Hamming distance}

Damage spreading simulations have become an important tool for exploring
time-dependent phenomena in statistical physics\cite{grasj}. By applying
it to spin systems or stochastic cellular automata (SCA) one follows the
evolution of a single difference (damage) between two replicas, driven by
the same random sequence. The initial damage may grow or shrink depending
on control parameters (temperature, etc.) of the model. If there is a phase
transition between such  regimes one can measure dynamical critical
exponents very precisely\cite{grasp}.

An important measure of the damage is the density of damaged sites.
In simulations one usually starts with a couple of systems ($N$) differing from
each other by a single variable, and  letting them evolve according to the
same rule and the same random sequence. For example in the case of a
two-state ($\{0;1\}$) system one measures the Hamming distance
between $N\times(N-1)/2$ pairs:
\begin{equation}
D(t) = \sum_{i=1}^{L^d} n_i(t) \, \left[N - n_i(t)\right]~, \label{hd}
\end{equation}
where $L$ is the system size in $d$ dimensions and $n_i(t)$ is the number
of `1'-s out of $N$ at site $i$. By tuning the control parameters the
measured $D(t)$ functions may follow the scaling law, and one can locate the
critical point and critical exponents by fitting.

Effective, multi-spin coded computer algorithms have been developed for serial
(i.e. DEC-ALPHA)\cite{grasp} and for parallel machines
(i.e. Connection Machine)\cite{grasdp}. The critical points and exponents of
phase transitions of different systems have been determined with great
accuracy\cite{grasj,grasp,grasdp}.

\subsection{ASP model}

We wrote a simple test program for the ASTRA machine to measure
the Hamming distances between $32\times 31/2$ pairs of lattices.
Simulating statistical mechanical systems on the ASP has been
discussed\cite{odo}. We have chosen Grassberger's cellular
automata model (A-model)\cite{grasa}. The program follows the evolution of
$N=32$ one-dimensional, two-state stochastic CA loaded along the ASP chain.
The calculation of $n_i$ and its two's complement $(32 - n_i)$ is executed
for all APE-s in parallel in a fixed number of steps ($128$ micro APE
instructions). The multiplication takes $7.5~\mu$s independently of $L$.
The summation for all processors can be done in a binary tree
algorithm, which requires ${\cal O}(\log L)$ summation steps. However, the
string communication of the present ASP machine allows only ${\cal O}(L)$
time scaling. This part of the calculation therefore dominates the execution
time until the necessary bypasses are built in the machine.
The total time of the Hamming distance calculation for $L=2^{14}$ systems is
$11.59$ ms and therefore the $73.5~\mu$s of the SCA update is negligible.
Here we assumed the present $20$ MHz clock speed and $16$K processors.
This means $707.4$ ns/site speed which compares with the $275$ ns/site value
that we achieved by the effective algorithm\cite{grasdp} on the Connection
Machine-5 of 512 processors, which should be qualified in view of the fact
that ASP applies extremely simple one-bit CPU-s compared to CM-5 with
SuperSPARC chips.

It should be that this algorithm could be run on the reduced ASP machine
proposed in Section~\ref{subsec:bitm}, which is specially configured for
addition in ${\cal O}(\log L)$ steps. Assuming $N=1024$ CA systems, and
$L=2^{20}$ processors, one can reach $4\times 10^{-9}~\mu$s per cell per pair
speed,
which corresponds to $10^4$ times the speed of a similar algorithm running on a

DEC-ALPHA machine\cite{grasp}.

\section{Pattern recognition}

It was proposed by A. Sandoval in 1988 in the framwork of the
LAA-project\cite{sando} to use the ASP massively parallel
system for pattern recognition in the NA35 heavy-ion experiment, where the
processing of one complete 3-D
event on three stereoviews took about 20 h in a semi-automatic
measuring device. In the meantime, by introducing   new powerful
RISC machines and more effective algorithms this problem was practically
solved, performing the trajectory recognition off-line for 200--300 particles
in a few minutes.

\subsection{Challenges in Pb--Pb physics}

However, in 1994 at the SPS,   the problem of complex event reconstruction
re-emerged
when  reconstructing several hundred  more
 particle trajectories for Pb--Pb
collisions. It will be even more challenging in the future at the LHC
when one will work with Pb--Pb colliding beams with TeV energies.
In the CERN-NA49 experiment the large Time-Projection-Chambers
produce about 8 MBytes/event
after zero suppression,   corresponding to a de-randomized
continuous data rate of 16 MBytes/s
which is   the present limit of maximal recording speed.
For the off-line analysis, a generalized 3-dimensional Hough-transform
was developed which needs 10 CPU seconds on a state-of-the-art RISC
computer (HP-755) to analyse 800 trajectories with about
70 space-points   each. For the LHC experiments, data rates
in the range of GigaBytes/s  are expected. There is therefore  plenty of room
for improvement in computer performance to reach real on-line
pattern recognition and to avoid recording mountains of redundant information.

Of course, the present ASTRA machine is too small to be competitive
in this market. Here we  only wish to demonstrate the perspectives
by a simple algorithmic example: what can be achieved by going to
really massive parallelism. In this case, although one strongly relies on the
associative search, i.e. on content addressing of the memory of
individual processors,  in practice one uses  less than 16-bit
comparators.
Consequently, most of the memory can still be realised by the standard memory
cells proposed in the above bit-matrix machine.

\subsection{Hough-transformation on binary image}\label{sec:hough}

For the simplicity of argument we assume that we have a (binary)
hit-list of $(x,y,z)$ coordinates, and we have to recognize straight line
tracks in a three-dimensional detector. We can specify tracks by $(x,y)$
coordinates of the first and last $z$ detector planes. To   balance
the complexity of the problem, we reduced it to a two-dimensional algorithm
by introducing a serial preselection in the $y$ direction.
We have now to determine the ($x_{\rm in}; x_{\rm out}$) data from the ($z,x$)
pairs
(Hough transformation) for each $y$ sector separately. It  still remains
a very difficult problem to process on-line hundreds of points per $y$-slice
with a $1000\times 1000$ resolution.

By pre-loading each possible ($x_{\rm in}; x_{\rm out}$) coordinate  per
processor,
a histo\-gram can be built that counts the number of incoming points matching
a track.
Having built the histogram, the maxima to be searched  correspond
to the tracks. One can immediately see that these tasks (i.e. matching, maximum
search) fit well the associative, parallel processing architecture.

It should be noted that we do not need to stick to binary image processing. The
algorithm can be extended to count intensities instead of hits without
any difficulty. However, we will need  more on-processor counter memory
for the more accurate summation of  amplitudes. The interest of recording
amplitudes for each cluster space-point is because these
amplitudes are correlated to the ionization of the track particle
which is velocity-dependent.

\subsection{ASP restricted model realization}\label{sec:asp}

To demonstrate the method we wrote a program assuming a 4K ASP
architecture. In this way we can have a $64\times 64$ ($x_{\rm in}; x_{\rm
out}$)
look-up table. Restricted by the $64$-bit APE memory, we divided the $y$
direction into three sectors. Therefore each processor contains three
histogram counters, pre-loaded data ($x_{\rm in}; x_{\rm out}$),  and working
serial fields. The matching condition is :
\begin{equation}
  \label{match}
  x_{\rm predicted} = x_{\rm in} + {(x_{\rm out} - x_{\rm in}) \over (z_{\rm
out} - z_{\rm in})}\times
 z~.
\end{equation}
To optimize the algorithm we:
\begin{itemize}
\item  assumed $(z_{\rm out} - z_{\rm in}) = 128$ resolution that reduces the
division operation to bit shifting (a similar situation can be achieved
by appropriate choice of units);
\item pre-loaded $(x_{\rm out} - x_{\rm in})$ values;
\item avoided the multiplication by a sequence of summations (after all
we have to go through all z-layers one by one).
\end{itemize}

The calculation of $x_{\rm predicted}$ takes $2.8~\mu$s per $z$ layer, while
the only time critical step, the `matching and counting', requires only
$1.35~\mu$s for each incoming data. This means an order of $1$ MHz processing
rate, assuming the present $20$ MHz clock speed independently of the
length of the processor string.

The processing speed is practically identical with the data loading speed.
The number of processors is important `only' for the accuracy of the
tracking, i.e. one needs as many processors as  possible track
combinations. In the present  NA49 Main-TPC for fine-grained
Hough-transform, $1000 \times 1000$ [$x_{\rm in},x_{\rm out}$] bins and 400
$y$-bins
are required. In total it represents 400 million counter cells. This
can be carried out on the machine with one million processors, because
for a given $x$ projection all the $y$-bins are stored in the same processor.
In order to achieve the required on-line speed, one would need to  speed-up
the present ASTRA machine by a factor of 100. Relying on advanced technology
one can hope to increase the clock frequency from 20 to 200 MHz. The
second factor of 10 should be achieved  by  increasing   the
I/O bandwidth, loading parallely at least 10 space-points.

\section{Conclusions}

The present crisis of tightly connected parallel (mainly SIMD)
computing can be overcome only by radical changes in technology, and
by entering into really massive parallelism of millions of cheap
processors. We suggest that these goals may be achieved by
adapting advanced memory technology to produce chips containing
more than 10 000 processors.

With relatively reduced, but essential, content addressing, all the relevant
matrix operations can be realized on {\bf string} architecture
with the `same' speed as on the {\bf mesh} one.

\section*{Acknowledgments}
We acknowledge partial support from the Hungarian National Research Fund (OTKA)

(grant no. F-7240, T-3271) and from the NCSA for CM-5 access
(grant no. PHY930024N).

\end{document}